\documentclass[twocolumn,aps,prl,groupedaddress,amsmath,amssymb]{revtex4-1}
\usepackage{url}
\usepackage[colorlinks=true, linkcolor=blue,urlcolor=blue,anchorcolor=blue,citecolor=blue,bookmarksnumbered]{hyperref}
\usepackage{graphicx}
\usepackage{dcolumn}
\usepackage{bm}
\usepackage{verbatim}

\begin{document}
	
	\title{Unidirectional acoustic metamaterials based on nonadiabatic holonomic quantum transformations}
	\author{JinLei Wu$^{1\dagger}$}\author{Shuai Tang$^{1\dagger}$}\author{Yan Wang$^{1}$}\author{XiaoSai Wang$^{1}$}\author{JinXuan Han$^{1}$}\author{Cheng L\"u$^{1}$}\author{Jie Song$^{1\ast}$}\author{ShiLei Su$^{2\ast}$}\author{Yan Xia$^{3}$}\author{YongYuan Jiang$^{1\ast}$}
	\affiliation{$^{1}$School of Physics, Harbin Institute of Technology, Harbin 150001, China\\
		$^{2}$School of Physics and Microelectronics, Zhengzhou University, Zhengzhou 450001, China\\
		$^{3}$Department of Physics, Fuzhou University, Fuzhou 350002, China\\
		$^{\ast}$Corresponding authors (Jie~Song,~email:~jsong@hit.edu.cn;\\
		ShiLei~Su,~email:~slsu@zzu.edu.cn; YongYuan~Jiang,~email:~jiangyy@hit.edu.cn)\\
		$^{\dagger}$These authors contributed equally to this work~}
	
	\begin{abstract}
		\noindent Nonadiabatic holonomic quantum transformations~(NHQTs) have attracted wide attention and have been applied in many aspects of quantum computation, whereas related research is usually limited to the field of quantum physics. Here we bring NHQTs into constructing a unidirectional acoustic metamaterial~(UDAM) for shaping classical beams. The UDAM is made up of an array of three-waveguide couplers, where the propagation of acoustic waves mimics the evolution of NHQTs. The excellent agreement among analytical predictions, numerical simulations, and experimental measurements confirms the great applicability of NHQTs in acoustic metamaterial engineering. The present work extends research on NHQTs from quantum physics to the field of classical waves for designing metamaterials with simple structures and may pave a new way to design UDAMs that would be of potential applications in acoustic isolation, communication, and stealth.\\
		\\
		\noindent{\bf acoustic metamaterial, waveguide coupler, holonomic quantum transformation\\
			\\
			PACS number(s):}~~~{43.40.+s, 78.67.Pt, 03.65.Vf, 42.50.-p}
	\end{abstract}
	\maketitle
	\noindent \textbf{\begin{large}1~~~{Introduction}\end{large}}\\
	\\
	With the development of acoustic metamaterials~(artificial materials made up of rationally designed building blocks), the last several decades have seen tremendous efforts dedicated to manipulating acoustic waves previously thought impossible~\cite{Cheng2021,Cummer2016}. Governed by generalized Snell’s law~\cite{Yu2011}, various acoustic devices, including wave concentrator~\cite{Hu2021}, beam splitter~\cite{Xie2017}, asymmetric absorber~\cite{Li2021}, and self-bending lens~\cite{Zhu2016}, were reported with one or more advantages of broadband, high-efficiency, and sub-wavelength features.
Acoustic unidirectional transmission has received a lot of attention as an interesting wave phenomenon because of its potential applications. For example, the acoustic diode can be effectively realized by breaking the time-reversal symmetry with nonlinear systems~\cite{Liang2009,Liang2010} or breaking the spatial inversion with an asymmetric geometry profile~\cite{Li2011}, and the tunability can be improved by a gradient metasurface with judiciously tailored loss~\cite{Li2017PRL}. Meanwhile, one-way mode conversion~\cite{Zhu2020} and non-Hermitian acoustic systems~\cite{Liu2020,Wang2019} have been demonstrated as good candidates for achieving unidirectional transmission. Research on acoustic unidirectional devices has advanced significantly from manipulating energy transport to exploring new physics connecting the asymmetric transmission to other intriguing modulations, including unidirectional acoustic invisibility~\cite{Zhu2014}, asymmetric focusing of sound~\cite{Liu2018}, and one-way sound localization~\cite{Shen2019}.

	There are also achievements connecting classical waves to topological and quantum effects~\cite{Shen2019,Shen2020,YXShen2020,Zeng2021,Chen2021PRL,Hu2021,Qiu2021,Li2021}, which may not only provide broader platforms for expanding categories of acoustic metamaterials but may also open up a new avenue for the construction of  functional metamaterials using previously unexplored approaches. We intend to build a bridge between unidirectional acoustic metamaterials~(UDAMs) and quantum computation. In the field of quantum computation~\cite{Knill2010,Preskill2021}, holonomic quantum computation~\cite{Zanardi1999} based on non-Abelian geometric phases~\cite{Berry1984,Wilczek1984} received substantial attention in the last two decades because the geometric evolution of quantum systems is not dependent on dynamical details but rather evolution trajectories and thus improves the fault tolerance of quantum computing~\cite{Duan2001,Zhu2002,Liu2019,TChen2020,ChenT2020,Long2021,PZZhao2021,Setiawan2021,Zhang2021,BJLiu2021}. Holonomic quantum computation was first designated based on adiabatic slow evolution~\cite{Zanardi1999,Duan2001,Wu2005}, and furthermore generalized to nonadiabatic paradigms~\cite{Sjoqvist2012,Xu2012,Feng2013,Abdumalikov2013,Zu2014,Arroyo_Camejo2014,Sekiguchi2017,Li2017,Zhou2017,GFXu2017,Zhou2018,Xu2018,GFXu2018,Yan2019,Ramberg2019,BJLiu2020,chen2021generation,Wu2021ARXIV,GFXu2021,PShen2021,SLi2021} that can efficiently shorten the time of holonomic transformations on quantum states and thus strengthen immunity to decoherence. However, using holonomic quantum transformations to build metamatierial devices is a relatively unexplored territory.
	
	\begin{figure}[b]\centering
		\includegraphics[width=0.98\linewidth]{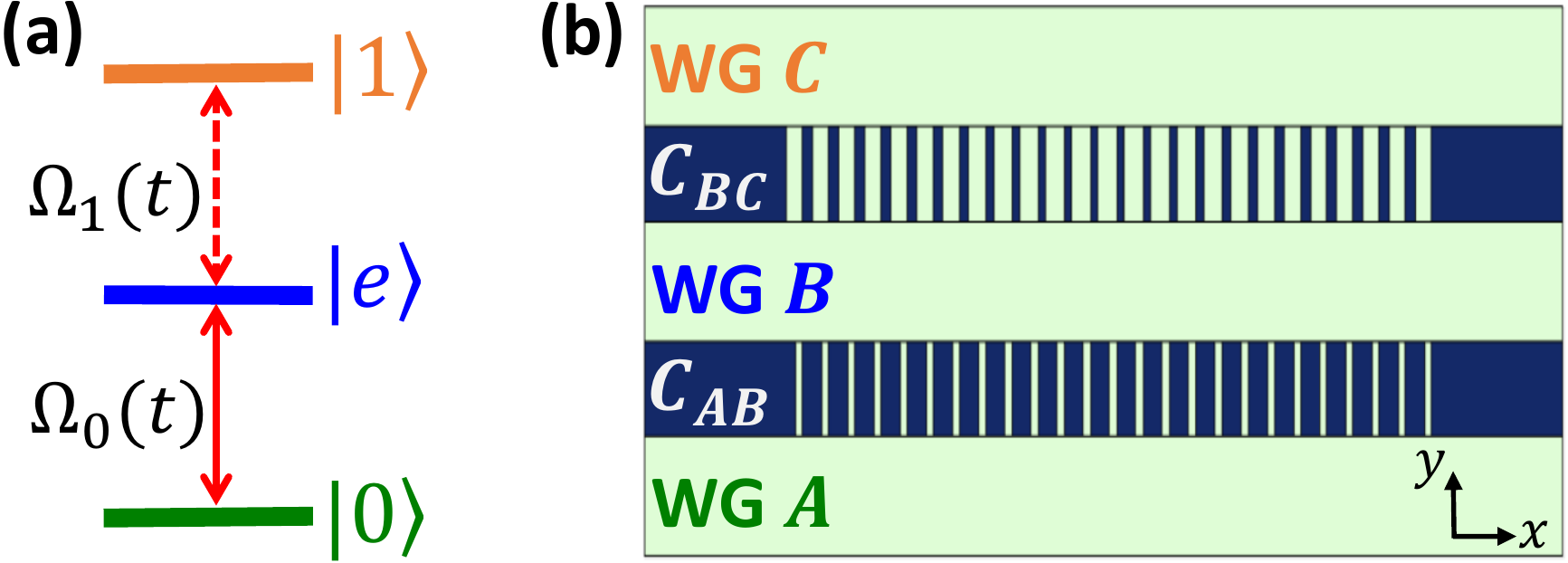}
		\caption{(a)~Three-level system with qubit states $|0\rangle$ and $|1\rangle$ coupled to an auxiliary state $|e\rangle$ with time-varying complex Rabi frequencies $\Omega_0(t)$ and $\Omega_1(t)$, respectively. (b)~Acoustic coupled-WG system with systematic WGs $A$ and $C$ getting coupled to an intermediate WG $B$ with space-dependent strengths $C_{AB}(x)$ and $C_{BC}(x)$. Acoustic energy transports between WGs are accomplished via  arrays of slits between WGs.}\label{f1}
	\end{figure}
	Existing methods for fabricating the acoustic meta-devices are typically associated with the fabrication of local resonators~\cite{Li2017PRL}, space-coiling units~\cite{Li2012}, or metagratings~\cite{Fu2019}, resulting in unavoidable structure design complexity. By introducing NHQTs into constructing acoustic metamaterial, this work provides a new way to guide the design of advanced acoustic functional devices from behind physical laws. In addition, the structural design only needs to be slotted periodically, making it easier for sample fabrication.	
		Meanwhile, due to the same distribution on both sides of the structure, it is difficult to realize unidirectional acoustic transmission and beam shaping simultaneously using conventional methods. Although bilayer design is a common approach for realizing asymmetric acoustic beam shaping~\cite{JPXia2018,YYFu2020}, while it is difficult to integrate. By taking advantage of the Hadamard-like sound transformation, our work provides a solution to achieve unidirectional beam splitting with a monolayer configuration, which may largely improve the practicability of the device.

	Assisted by agreement in form between Schr\"odinger equation in quantum mechanics and the coupled-mode equation of classical waves, coupled waveguides~(WGs) for sound propagation can be mapped to quantum states driven by external fields, allowing NHQTs on discrete states in a three-level quantum system~[see Fig.~\ref{f1}(a)] to be used to engineer energy transfer in an acoustic three-WG coupler~[see Fig.~\ref{f1}(b)]. Here we focus mainly on designing a beam splitter by mimicking a Hadamard transformation on the WG coupler, and also we show that a directional coupler of exchanging acoustic energy can be designed directly to mimic a NOT gate in quantum information processing, or indirectly by combining two beam splitters with a device of the phase shifter. Furthermore, we design and fabricate an acoustic metamaterial capable of efficiently splitting a plane acoustic wave in a relatively large space by arraying several beam splitters. In addition, the designed metamaterial device is unidirectional, which means that the plane acoustic wave is split when input from one side but barely transmitted when input from the other. We identify the excellent agreement among results of analytical predictions, numerical simulations, and experimental measurements for the intensity distribution of the acoustic field transmitting through the UDAM, which indicates that the present work provides a practical approach for designs of metamaterials.\\
	\\
	\noindent \textbf{\begin{large}2~~~{Model and method}\end{large}}\\
	\\
	Because of the mathematical similarity between the time-dependent Schr\"odinger equation and the wave equation describing the spatial propagation of the classical wave, the application of quantum control techniques, such as adiabatic passage~\cite{Bergmann1998,Kral2007,Vitanov2017}, in systems of coupled WGs has grown rapidly in recent years~\cite{Paspalakis2006,Hristova2016,Shen2019,JChen2021}. On the one hand, propagating properties of classical waves in WGs provide an intuitional visualization in the space of typical ultrafast phenomena in time. WGs of classical waves, on the other hand, are excellent platforms for exploring coherent dynamical regimes that are difficult to access in quantum systems. We then consider an analogy between the classical-wave coupled-mode equations and the quantum Schr\"odinger equation, in which intensities and complex-valued acoustic pressures of sound propagating in WGs, respectively, play the roles of probabilities and probability amplitudes of wave functions in quantum mechanics.

	We start by considering a model of $\Xi$-type three-level system in the context of quantum computation, as shown in Fig.~\ref{f1}(a). The transition between a qubit state $|0(1)\rangle$ and the auxiliary state $|e\rangle$ is achieved through pumping by external fields. Under the rotating-wave approximation the interaction-picture Hamiltonian of this three-level system can be written as~($\hbar=1$ hereinafter) $\hat H_i(t)=\sum_{j=0}^1\Omega_j(t)|e\rangle\langle j|+\Omega_j^\ast(t)|j\rangle\langle e|$ with the superscript ``$\ast$" denoting the complex conjugation. For the system initially in the qubit state $|k\rangle$~($k=0,1$), the state of the system at each instant $t$ can be obtained as $|\psi_k(t)\rangle=\mathcal{\hat U}(t)|k\rangle$ with the evolution operator $\mathcal{\hat U}(t)=\mathcal{\hat T}\exp(-{\rm i}\int_0^t\hat H_i(t')dt')$ according to Schr\"odinger equation ${\rm i}\partial |\psi_k(t)\rangle/\partial t=\hat H_i(t)|\psi_k(t)\rangle$, $\mathcal{\hat T}$ denoting time ordering.
	
	\begin{figure}[b]\centering
		\includegraphics[width=0.8\linewidth]{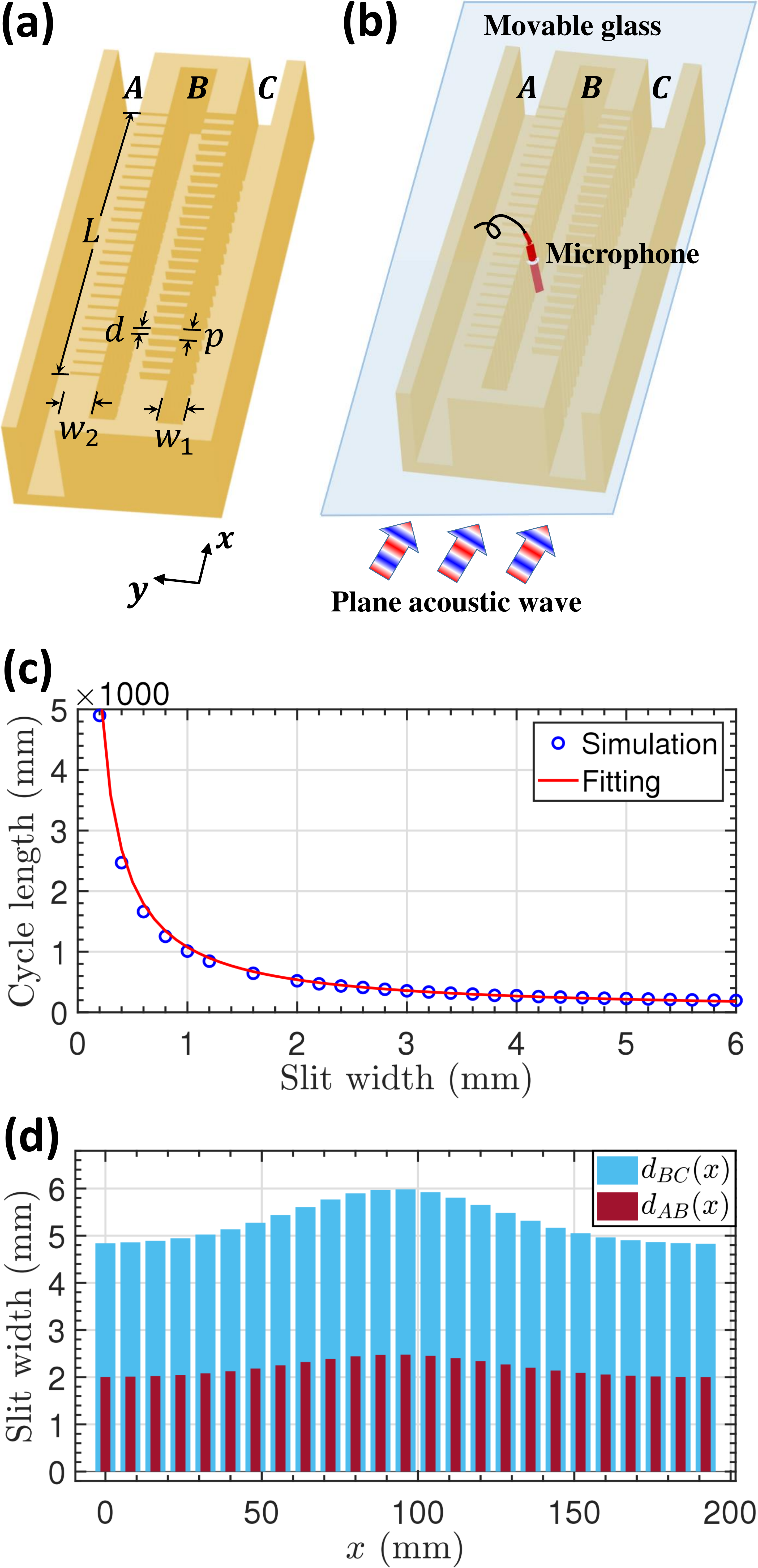}
		\caption{Sample diagram of the three-WG coupler. The plane acoustic wave is normally incident from (a)~solely WG $A$ or (b)~homogeneously WGs $A$ and $C$. A microphone was used to measure sound pressure at each site of the three WGs. The microphone is mounted on a movable piece of glass that completely encloses the sample. The experimental samples in (a) and (b) are taken from the UDAM sample, which is depicted in Fig.~\ref{f5}(c). The work frequency is $4400$~Hz.}\label{f2}
	\end{figure}
	\begin{figure*}\centering
		\includegraphics[width=0.88\linewidth]{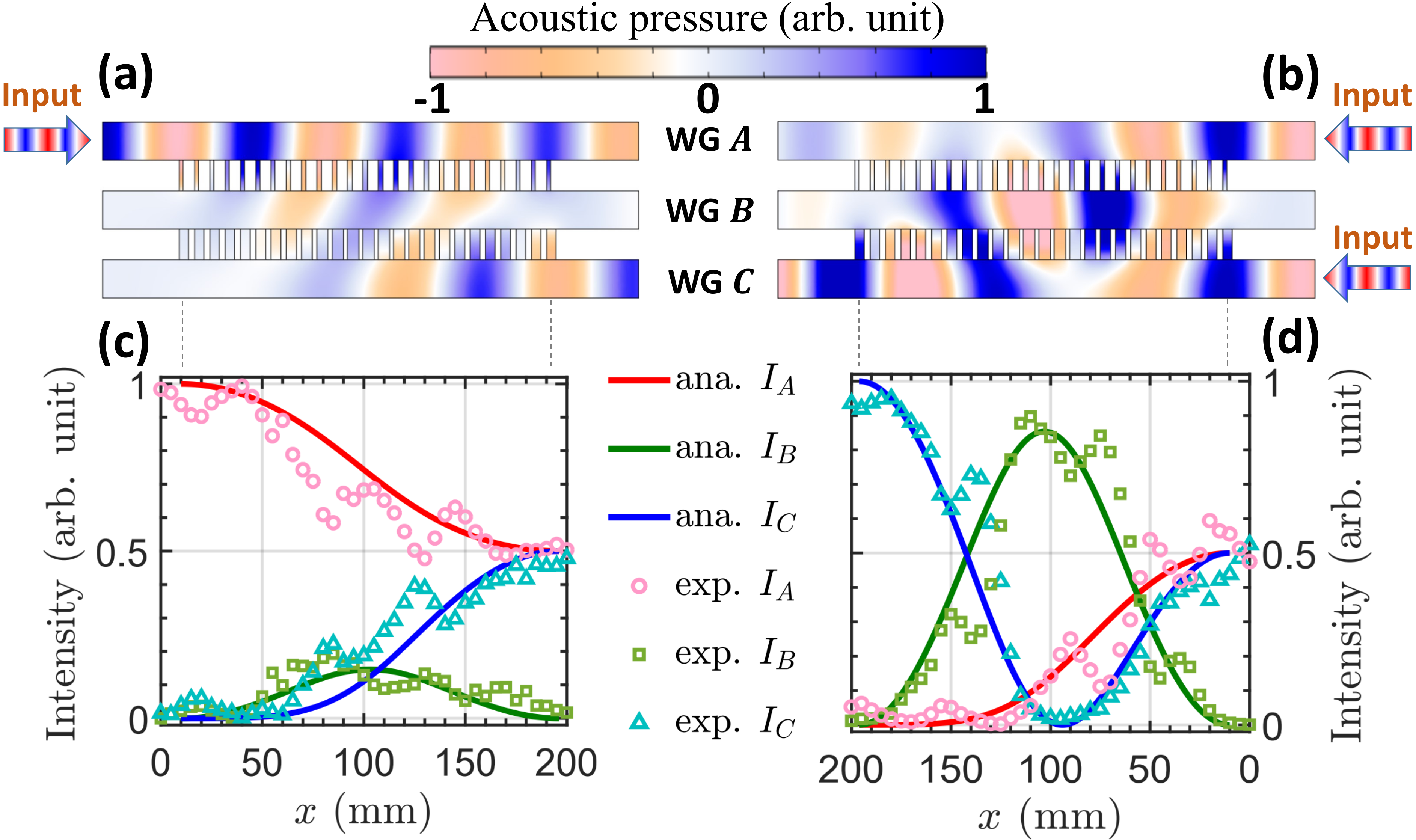}
		\caption{Simulation for acoustic pressures in the three-WG coupler with the plane acoustic wave normally incident from (a)~solely WG $A$ or (b)~homogeneously WGs $A$ and $C$. In contrast to analytic predictions, experimental results for acoustic intensities in the three-WG coupler with the plane acoustic wave normally incident from (c)~solely WG $A$ or (d)~homogeneously WGs $A$ and $C$.}\label{f3}
	\end{figure*}
	Following the method of implementing nonadiabatic holonomic quantum gates in three-level systems~\cite{Sjoqvist2012,Xu2012,Feng2013,Abdumalikov2013,Zu2014,Arroyo_Camejo2014,Sekiguchi2017,Li2017,Zhou2017,Zhou2018,Xu2018,Yan2019}, the two Rabi frequencies can be parameterized by a time-independent mixed angle $\theta$, a relative phase $\phi$, and an amplitude $\Omega(t)$ through relations $e^{{\rm i}\phi}\tan(\theta/2)=\Omega_0(t)/\Omega_1(t)$ and $\Omega(t)=\sqrt{|\Omega_0(t)|^2+|\Omega_1(t)|^2}$. Then two dressed states are formed, $|b\rangle=e^{-{\rm i}\phi}\sin(\theta/2)|0\rangle+\cos(\theta/2)|1\rangle$ and $|d\rangle=e^{-{\rm i}\phi}\cos(\theta/2)|0\rangle-\sin(\theta/2)|1\rangle$, where the former is coupled to the auxiliary state $|e\rangle$ with strength $\Omega(t)$ while the latter decouples from the dynamics. For implementing nonadiabatic holonomic gates on qubit states \{$|0\rangle$,~$|1\rangle$\}, two conditions are supposed to be satisfied (i)~the cyclic evolution condition $\sum_{k=0,1}|\psi_k(T)\rangle\langle\psi_k(T)|=\sum_{k=0,1}|\psi_k(0)\rangle\langle\psi_k(0)|$ with $T$ being the ending time of the evolution; (ii)~the parallel-transport condition $\langle\psi_k(t)|\hat H_i(t)|\psi_l(t)\rangle=0$ with $k,l=0,1$~\cite{Sjoqvist2012,Xu2012}. The former can be satisfied by setting the pulse duration $\int_0^T\Omega(t)dt=\pi$ such that $\mathcal{\hat U}(T)=|d\rangle\langle d|-|b\rangle\langle b|$. The latter is ensured by setting a time-independent $\theta$, and thus the accumulated dynamical phase during the entire evolution of the system is null, resulting in a purely geometric $\pi$ phase on $|b\rangle$ purely geometric. With the basis of \{$|0\rangle$,~$|1\rangle$\}, the resulting evolution operator is of the form
	\begin{equation}\label{e1}
		\mathcal{\hat U}(T)=\left(
		\begin{array}{cc}
			\cos\theta&-e^{-{\rm i}\phi}\sin\theta\\
			-e^{-{\rm i}\phi}\sin\theta&-\cos\theta\\
		\end{array} \right).
	\end{equation}
	This is a holonomic gate that can be used to conduct transformations on a single qubit, such as the following three familiar single-qubit gates~(containing global phases)
		\begin{equation*}
			\sigma_x=\left(
			\begin{array}{cc}
				0&-1\\
				-1&0\\
			\end{array} \right),~
			\sigma_z=\left(
			\begin{array}{cc}
				1&0\\
				0&-1\\
			\end{array} \right),~
			{\rm H}=\left(
			\begin{array}{cc}
				\frac1{\sqrt2}&\frac{-1}{\sqrt2}\\
				\frac{-1}{\sqrt2}&\frac{-1}{\sqrt2}\\
			\end{array} \right),
		\end{equation*}
		by setting $\{\theta=\pi/2,~\phi=0\}$, $\{\theta=0,~\phi=0\}$, and $\{\theta=\pi/4,~\phi=0\}$, respectively, called as NOT, $\pi$-phase, and Hadamard gates. In addition, other transformations can be obtained  by setting suitable parameters $\{\theta,~\phi\}$ or combining two easily accessible gates as needed~\cite{Sjoqvist2012}.

	In the acoustic coupled-WG system, the three WGs $A$, $B$, and $C$ play the roles of the discrete quantum states $|0\rangle$, $|e\rangle$, and $|1\rangle$, respectively. By mechanically opening discrete slits between two adjacent WGs, as shown in Fig.~\ref{f1}(b), these two WGs can be coupled by transporting sound energy, for which the coupling strength can be space-dependent by varying the slit widths along the $x$-axis. Concurrently, the coupling strength is also time-dependent because time and space are connected by  the sound velocity. Therefore, the inter-WG coupling strengths $C_{AB}(x)$ and $C_{BC}(x)$ are corresponding, respectively, to the Rabi frequencies $\Omega_{0}(t)$ and $\Omega_{1}(t)$ of transitions between quantum states. The propagation of the acoustic field in the coupled WG structure is described by the coupled-mode theory~\cite{Messiah1962}
	\begin{equation}\label{e2}
		\begin{split}
			&{\rm i}\frac{\partial P_A(x)}{\partial x}={C_{AB}}^\ast(x)P_B(x),\\
			&{\rm i}\frac{\partial P_B(x)}{\partial x}={C_{AB}}(x)P_A(x)+C_{BC}(x)P_C(x),\\
			&{\rm i}\frac{\partial P_C(x)}{\partial x}={C_{BC}}^\ast(x)P_B(x),
		\end{split}
	\end{equation}
	where we have defined $P_l(x)$~[$|P_l(x)|^2$] as the acoustic pressure~(intensity) in WG $l$~($l=A,B,C$), and normalization $|P_A(x)|^2+|P_B(x)|^2+|P_C(x)|^2=1$ is taken into account. Here we use the complex inter-WG coupling strength ${C_{lm}}(x)$~($lm=AB,BC$), for which $-{\rm i}\ln[{C_{lm}}(x)/|{C_{lm}}(x)|]$ denotes the phase difference between WGs. In practice for our model, this phase difference can be achieved by installing a phase shifter in one WG~\cite{Li2012,Xie2014}.

	It is straightforward to write Eq.~(\ref{e2}) in the form similar to Schr\"odinger equation in the space dimension
	\begin{equation}\label{e3}
		{\rm i}\frac{\partial |\Psi(x)\rangle}{\partial x}= \hat M(x)|\Psi(x)\rangle.
	\end{equation}
	The coupling operator of the three WGs corresponds to the Hamiltonian of the quantum three-level system, $\hat M(x)=C_{AB}(x)|B\rangle\langle A|+C_{BC}(x)|B\rangle\langle C|+{\rm h.c.}$, with ${\rm h.c.}$ denoting Hermitian conjugation, for which we have introduced acoustic states for the three WGs as the basis, $|A\rangle$, $|B\rangle$, and $|C\rangle$. The state~(sound pressure) of the three-WG system at each site, $|\Psi(x)\rangle=P_A(x)|A\rangle+P_B(x)|B\rangle+P_C(x)|C\rangle$, can be obtained by solving the Schr\"odinger-like equation~[Eq.~(\ref{e3})]. Then  when we set a constant ratio $C_{AB}(x)/C_{BC}(x)=\tan(\theta/2)e^{{\rm i}\phi}$ and also $\int_0^Ldx\sqrt{|C_{AB}(x)|^2+|C_{BC}(x)|^2}=\pi$ with $L$ being the WG length of coupling, the acoustic transformation between WGs $A$ and $C$ from the input ends to the output ends follows the NHQT in Eq.~(\ref{e1}). Consequently, by designing arrays of slits between WGs we can fabricate acoustic couplers with some specific functions, such as mode transport between WGs $A$ and $C$~($\theta=\pi/2$), selective phase flip of a certain WG~($\theta=0$), beam splitting with an arbitrary proportion~($\phi=0$) between WGs $A$ and $C$.\\
	\\
	
	\begin{figure*}\centering
		\includegraphics[width=0.88\linewidth]{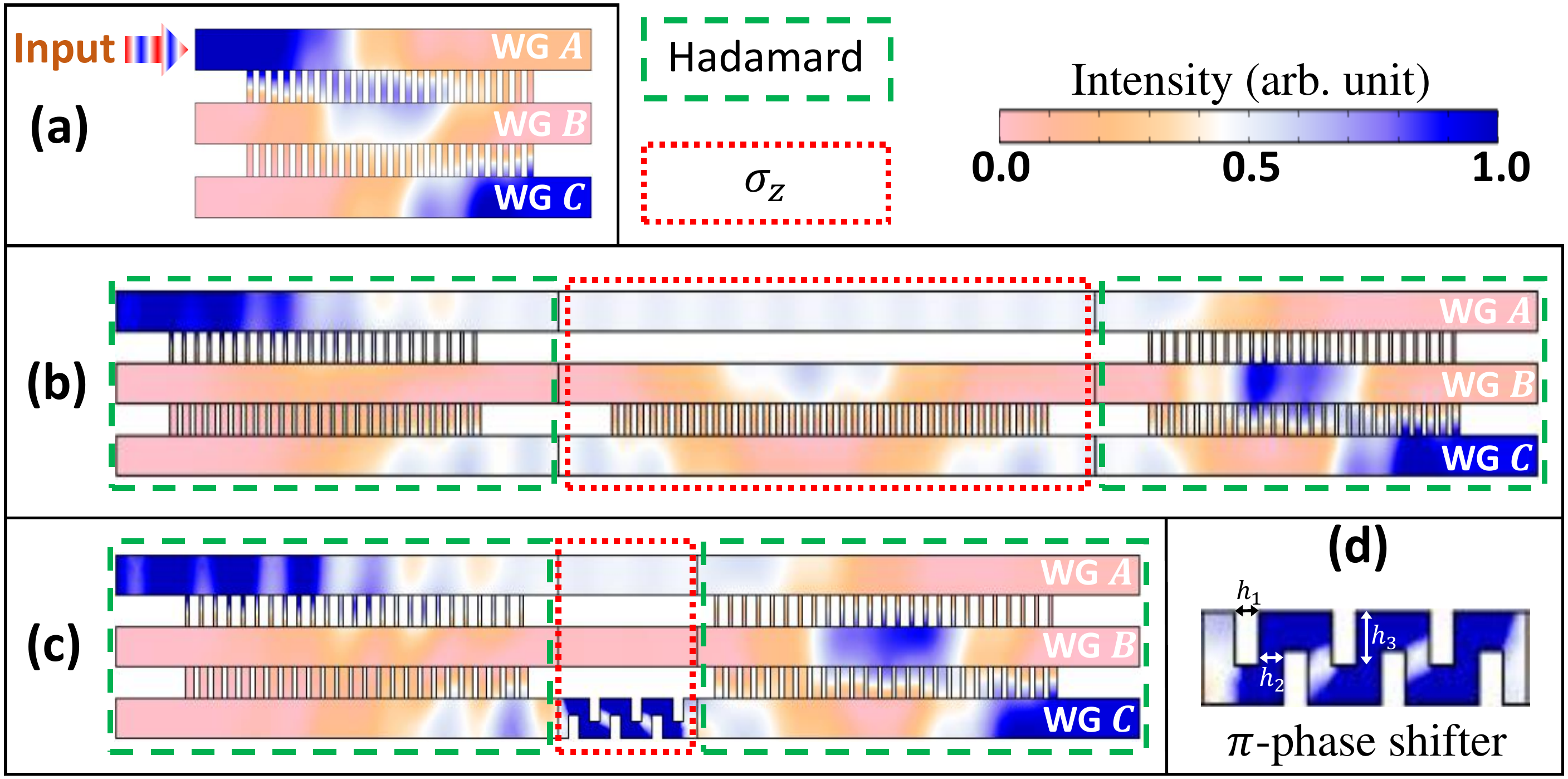}
		\caption{Stimulated implementations of a NOT sound transformation on three WGs by (a)~directly coupling three WGs, or combining two Hadamard gates before and after a $\sigma_z$ transformation, where the $\sigma_z$ gate is realized by (b)~coupling WGs $B$ and $C$ or (c)~installing a $\pi$-phaee shifter on WG $C$. (d)~Configuration of the $\pi$-phaee shifter configured with $h_1=h_2=3$~mm and $h_3=5.8$~mm.}\label{f4}
	\end{figure*}
	\noindent \textbf{\begin{large}3~~~{Acoustic three-WG beam splitter}\end{large}}\\
	\\
	We design and fabricate an acoustic three-WG coupler that renders a Hadamard-like sound transformation~($\theta=\pi/4$ and $\phi=0$) from the input ends to the output ends
	\begin{equation}\label{e4}
		|A\rangle\rightarrow\frac1{\sqrt2}(|A\rangle-|C\rangle),\quad|C\rangle\rightarrow-\frac1{\sqrt2}(|A\rangle+|C\rangle),
	\end{equation}
	where the phase of the acoustic pressure in each WG is relative to that of a free WG. This acoustic coupler, is used to create a beam splitter. Acoustic intensities in WGs $A$ and $C$ are identical,  at the output ends, while the phases of the acoustic waves are opposite~(identical) when sound is incident from WG $A$~($C$).

	The sample of the three-WG coupler with the function of beam splitting is schematized in Fig.~\ref{f2}(a)~[in Fig.~\ref{f2}(b)], where the plane acoustic wave is normally incident from solely WG $A$~(from homogeneously WGs $A$ and $C$). The sample is made of a photosensitive resin via stereolithography~(SLA, 0.2 mm in precision). The mass density, Young modulus, and Poisson ratio of the resin are $\rho_r=1050$~kg/m$^3$, $E_r=2.65$~GPa, and $v=0.41$, respectively. When compared to air, the resin is acoustically rigid. The three parallel acoustic WGs are filled with air connected by equally spaced slits with period $p=8$~mm. The width of each air pipe is $w_1=20$~mm, and the thickness between two adjacent air pipes is $w_2=16$~mm. The numerical relationship between the inter-WG coupling strength $\mathcal{C}$ and the slit width $d$ can be determined by $\mathcal{C}=\pi/l$~\cite{Shen2019}, where $l$ describes the length of a cycle for energy oscillating between two WGs and is closely related to $d$. According to the simulation results from the finite element method, we have calculated out data of $l$ versus constant $d$ for the sample of two coupled WGs, as shown in Fig.~\ref{f2}(c), which can be fitted well by the function $l=\alpha/d$ with $\alpha=1075~{\rm mm}^2$.

	Due to the limitation of fabrication craft, slit widths should be machined with $d,~(p-d)\geq 2~{\rm mm}$ to ensure adequate accuracy. On the other hand, to average out effects of the fabrication imperfections and the $l$-$d$ fitting errors, we apply a space-dependent slit array $d_{AB}(x)$~[$d_{BC}(x)$] between WGs $A$ and $B$~[$B$ and $C$], following the conception of Gaussian soft quantum control~\cite{Haase2018}:
	\begin{equation}\label{e5}
		\begin{split}
			&d_{AB}(x)=\tan(\theta/2)C_{BC}(x),\\
			&d_{BC}(x)=d_0\exp[-(x-x_0)^2/\chi^2]+\eta d_0,
		\end{split}
	\end{equation}
	where $\theta=\pi/4$, $x_0=L/2$ and $\chi=L/4$. In addition, we also consider $\min[d_{AB}(x)]=2$~mm, $\max[d_{BC}(x)]=6$~mm, and $\int_0^Ldx\sqrt{|C_{AB}(x)|^2+|C_{BC}(x)|^2}=\pi$, and then we determine $d_0=1.17$~mm, $\eta=4.12$, and $L=192$~mm. Based on parameters above, the slit arrays with space-dependent widths $d_{AB}(x)$ and $d_{BC}(x)$ are designed as shown in Fig.~\ref{f2}(d), each array containing 25 holes.  
	
	\begin{figure*}\centering
		\includegraphics[width=0.7\linewidth]{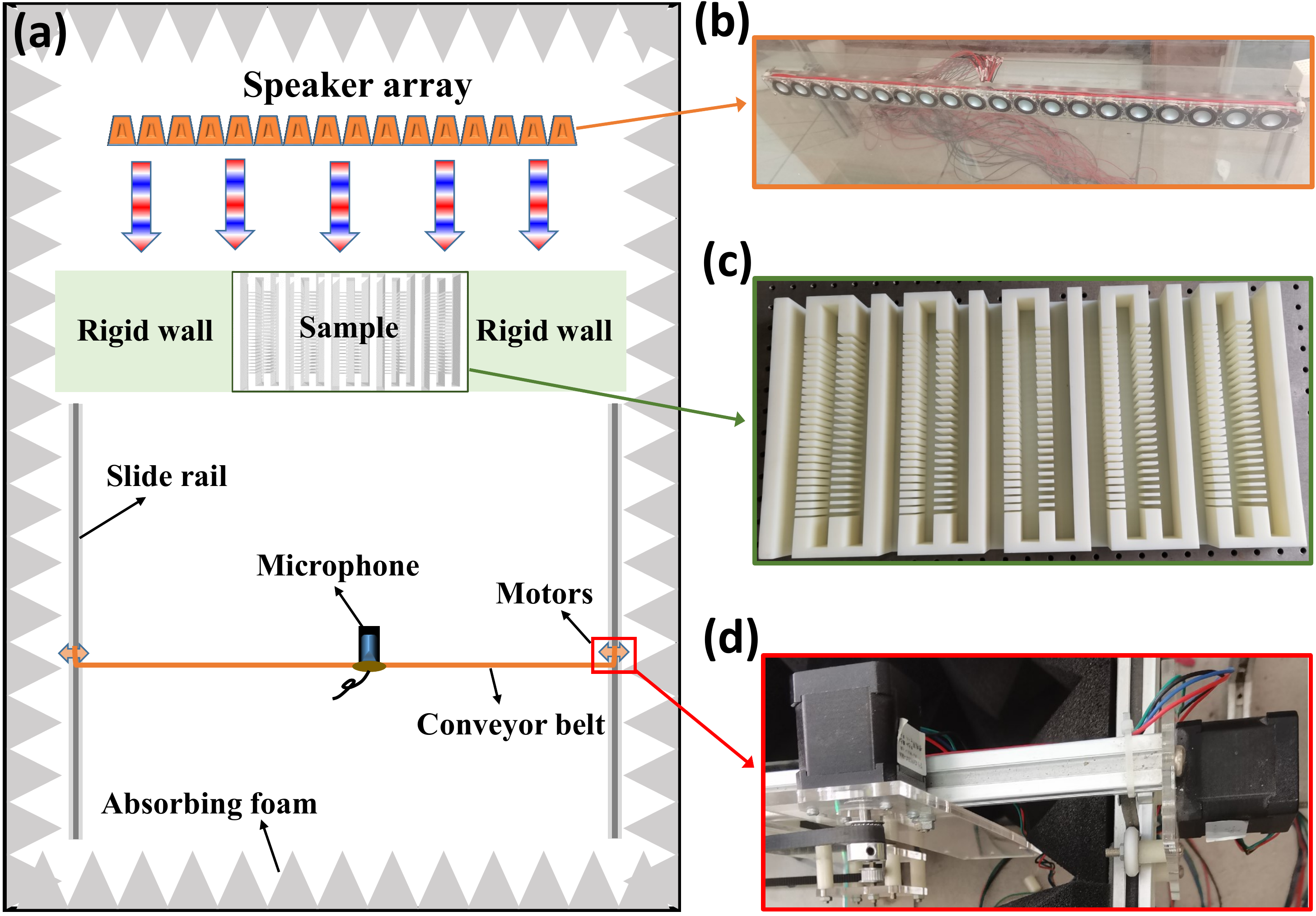}
		\caption{(a)~Experimental platform for measuring acoustic intensity distribution behind the UDAM. Photographs of (b)~the arrayed loudspeakers, (c)~the sample for the UDAM, and (d)~stepper motors for moving the microphone.}\label{f5}
	\end{figure*}
	According to the analytical result in Eq.~(\ref{e4}), the acoustic wave should emit from WGs $A$ and $C$ with identical intensities but opposite phases for the three-WG sample in Fig.~\ref{f2}(a), but only from WG $C$ for that in Fig.~\ref{f2}(b). In Figs.~\ref{f3}(a) and (b), we use the finite element method to simulate the propagation of acoustic pressure in the couplers, and the simulations agree very well with the analytical predictions in Eq.~(\ref{e4}). Furthermore, we experimentally measure the acoustic intensities at each site of the three WGs in the samples in Figs.~\ref{f2}(a) and (b), respectively. The experimental measurements are shown in Figs.~\ref{f3}(c) and (d), where we also give the analytic results by solving the Schr\"odinger-like equation~[Eq.~(\ref{e3})] for comparison. Except for some oscillation phenomena caused by interference between incident and slight reflections of acoustic waves, the experimental results are nearly identical to the analytic predictions. The simulative and experimental results identify the practicability of the NHQT in designing acoustic WG couplers.
	
	In addition to the Hadamard sound transformation, other types of single-qubit gates can also be mimicked by the three coupled WGs. In Fig.~\ref{f4}, we show results of simulating three methods of implementing a NOT gate that exchanges the sound energies in WGs $A$ and $C$. Figure~\ref{f4}(a) verifies the direct construction of the NOT transformation by setting identical and constant coupling slit widths between two pairs of adjacent WGs to ensure~$\{\theta=\pi/2,~\phi=0\}$, where each array has 25 coupling holes and the slit width is $4$~mm. In addition to the direct construction, a NOT gate can be obtained indirectly by combining two Hadamard gates before and after a $\pi$-phase gate according to the equation~$\sigma_x={\rm H}\cdot\sigma_z\cdot {\rm H}$. Referring to Eq.~(\ref{e1}), $\sigma_z$ is achieved with $\{\theta=0,~\phi=0\}$, demanding no coupling between WGs $A$ and $B$, so a NOT gate can be achieved as the simulation in Fig.~\ref{f4}(b), where between WGs $B$ and $C$ the number of coupling holes is 35 and the slit width is $4$~mm for realizing the $\pi$-phase shift of sound wave in WG $C$. Alternatively, a $\pi$-phase shift of a sound wave in WG $C$ can be obtained by installing a phase shifter on it. By coiling up space in the WG~\cite{Li2012}, an arbitrary phase shift can be achieved flexibly. Therefore, a NOT sound transformation can be simulated by installing a $\pi$-phase shifter on WG~$C$ between two Hadamard gates, as shown in Fig.~\ref{f4}(c), for which the configuration of this $\pi$-phase shifter is exhibited in Fig.~\ref{f4}(d). Other shifters of arbitrary phases can also be built in this 
	manner to shift the phase difference between two WGs and thus adjust the parameter $\phi$, and also using a phase shifter can also shorten the length of the device~[see Figs.~\ref{f4}(b) and (c)].\\
	\\
	
	\noindent \textbf{\begin{large}4~~{Unidirectional acoustic metamaterial}\end{large}}\\
	\\
	We adopt a UDAM composed of five~(or more) WG couplers illustrated in Fig.~\ref{f2}(a) to achieve an asymmetric beam splitter at 4400~Hz. Figures~\ref{f5}(a) and (c) show the experimental platform and a sample photograph for the UDAM. The experimental setup is based on a planar waveguide system, with the sample sandwiched between two pieces of tempered glass. To eliminate unwanted sound reflections from the outside, absorbing foams are wrapped around the system. As illustrated in Fig.~\ref{f5}(b), 20 tightly arrayed loudspeakers are adopted as an acoustic source to generate the incident plane acoustic waves. To detect sound signals, a mechanically movable microphone is driven by stepper motors~[Fig.~\ref{f5}(d)]. The transmitted field at any point behind the sample can be easily scanned by moving the microphone along the slide rail and conveyor belt, allowing the intensity distribution feature of the acoustic field to be obtained invertedly.
	
	\begin{figure*}\centering
		\includegraphics[width=0.88\linewidth]{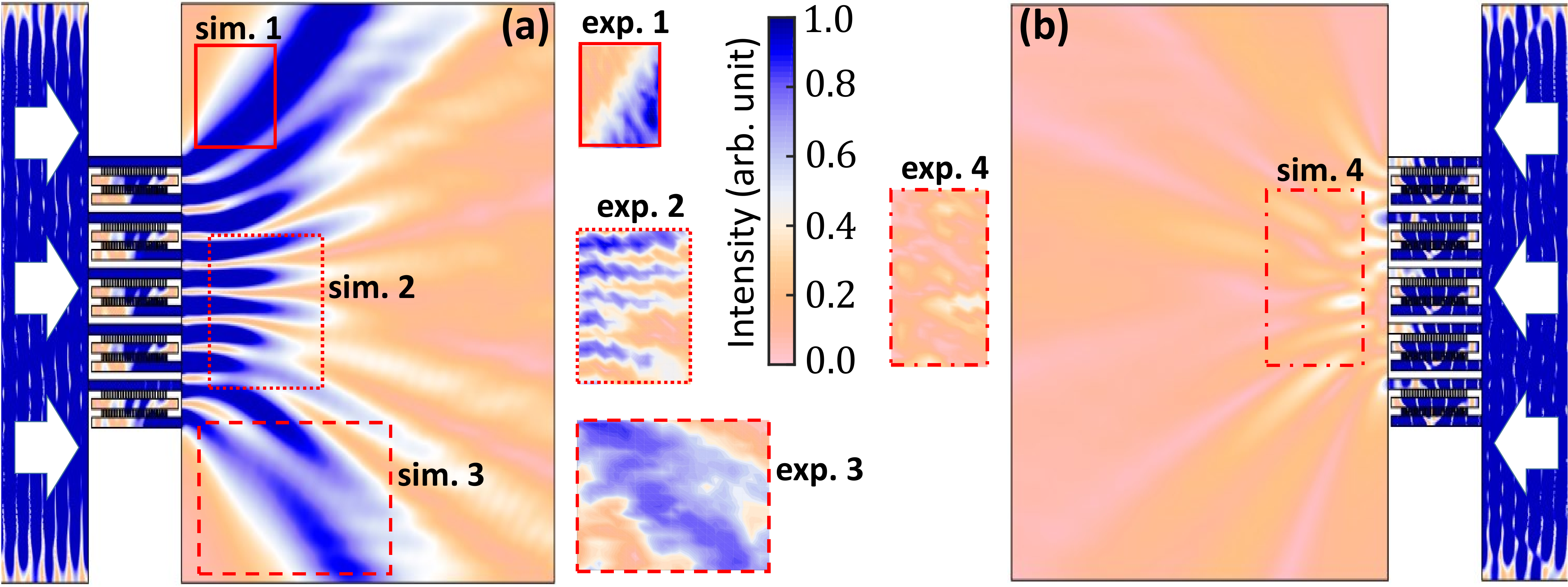}
		\caption{Comparison of simulation and experimental results for the measured intensity distribution of the acoustic fields behind a three-WG coupler array with the plane acoustic wave incident from (a)~solely WG $A$ or (b)~homogeneously WGs $A$ and $C$.}\label{f6}
	\end{figure*}
	It has been demonstrated in Fig.~\ref{f3}(a) that the transmission phases are opposite at the end of WGs $A$ and $C$ when plane acoustic waves are incident from WG $A$. A beam splitter can be built using the $\pi$-phase difference between the two WGs. The acoustic field pattern of a period structure composed of two types of unit cells with opposite transmitted phases is determined by the period length~$P_L$ and the incident wavelength~$\lambda$, according to the generalized Snell’s law~\cite{Yu2011}. When $P_L<\lambda$, the value of reciprocal lattice vector $G=2\pi/P_L$ is larger than the wave number $k_0$. The diffracted waves of non-zero orders are evanescent, and the acoustic splitting is not able to be obtained. When $P_L>\lambda$ and $G<k_0$, diffraction of the $0$-order, $+1$-order, and $-1$-order are all allowed. The diffraction of the $+1$-order and the $-1$-order are able to directly take one-pass propagation, which is preferential to the $0$-order, according to the round-trip process demonstrated in the work~\cite{Fu2019}. Thus, the beam splitting is capable of achieving and the refraction angle can be calculated by $\theta_r=\arcsin(\lambda/P_L)$. In the present work, the operating wavelength is $\lambda=(343~{\rm m/s})/(4400~{\rm Hz})=0.078~{\rm m}$. We set $P_L=0.092~{\rm m}$, which is longer than the wavelength, to avoid the generation of the evanescent wave. The plane acoustic wave normally incident from the left side of the metamaterial is split into two different directions with $\theta_r=1$~rad, as shown in Fig.~\ref{f6}(a). The insets of Fig.~\ref{f6}(a) show the measured intensity fields corresponding to the red boxes marked in the simulated field. Excellent agreements between simulation and experimental results confirm the presence of beam splitting features in the transmission field.

	However, as shown in Fig.~\ref{f3}(b), high-intensity energy only propagates along the end of WG $C$ when the plane acoustic wave is incident from ports of WGs $A$ and $C$ at the same time, providing a solution to achieve unidirectional transmission feature of the coupled-WG system by sealing the output of WG $C$ with a rigid wall. As a result, it is indicated in Fig.~\ref{f6}(b) that the transmitted intensity of the acoustic field is quite weak when the plane wave is normally incident from right side of the metamaterial, as shown in Fig.~\ref{f6}(b). The measured intensity distribution of the transmitted field illustrated in the experimental result~(exp.~4) is almost coincident with that in the simulated field~(sim.~4). The intensity of the transmitted field for the right-side incidence is relatively weak when compared to the left-side incidence due to the presence of strong reflection from the output of WG $C$ caused by the sealing rigid wall. It exactly verifies important features of unidirectional transmission and beam splitting for the proposed metamaterial, which is built with NHQT-designed three-WG couplers.
	
	Given the fact that nonadiabatic holonomic quantum computation~(NHQC) has been widely investigated and there are remarkable achievements for implementing fast and robust quantum gates~\cite{Feng2013,Abdumalikov2013,Zu2014,Arroyo_Camejo2014,Sekiguchi2017,Li2017,Zhou2017,Zhou2018,Xu2018,Yan2019,Wu2021ARXIV,chen2021generation}, on the one hand the present work would be a significant enrichment of these achievements and largely expand the realm of applying NHQC by bringing NHQTs into acoustic wave manipulation. On the other hand, the present work only demonstrates the use of a relatively simple type of NHQC in metamaterials applications. For further research, NHQC combined with specific control technologies, such as NHQC combined with optimal control~\cite{Liu2019,Wu2021ARXIV}, could greatly extend the excellent properties of metamaterials, such as broadband, compaction, and robustness against craft errors. Besides the beam-splitting metamaterial device identified above, other beam-shaping devices, beam self-accelerating~\cite{Zhu2016}, beam focusing~\cite{Li2012}, and beam steering~\cite{Xie2014}, could also be designed by NHQTs with different pairs of $\theta$ and $\phi$.\\
	\\
	\noindent \textbf{\begin{large}5~~~{Discussion and conclusion}\end{large}}\\
	\\
	The NHQTs considered above~[Eq.~(\ref{e1})] are achieved by the way that was reported first in Ref.~\cite{Sjoqvist2012}. There are also other types of nonadiabatic holonomic quantum gates, such as single-shot gates with off-resonant three-level structures~\cite{GFXu2015,Sjoqvist2016} and multipulse single-loop settings~\cite{Herterich2016,ZPHong2018}, where the former introduces a degree of freedom, i.e., detuning parameter, to enable single-shot implementations of arbitrary rotations of a single qubit, and the latter renders the gate trajectory on the Bloch sphere to be an ``orange-slice" path enclosed by two or more segments so that an arbitrary gate can be achieved by a single loop. These two forms of NHQTs may be more convenient to some extent to implement an arbitrary sound transformation between WGs than that used here. Exploring the realizations of these two types of NHQTs on acoustic three-WG systems is therefore of interest. To introduce a detuning parameter~\cite{GFXu2015,Sjoqvist2016}, one can couple WGs with different widths $w_1$ along the $y$-axis~[see Fig.~\ref{f2}(a)]~\cite{Shen2020}. For mimicking multipulse couplings with the pulses being of different phases, one can divide the coupled WGs into several coupling parts along $x$-axis, and adjacent parts are connected by phase shifters~\cite{Li2012} to adjust the phase difference between sound waves, which corresponds to the phase difference between different pairs of pulses~\cite{Herterich2016}.
	
	Regardless of the flexibility in achieving the UDAMs based on NHQTs in three-WG systems, for the first proposal of introducing NHQTs into designing acoustic metamaterials, there are also some limitations. First, the precision of discrete coupling holes should be taken into account when mapping the time-dependent driving pulses into spatial dimensions. As a result, the slit period length must be as short as possible to guarantee enough discrete points in a fixed device length. Due to the limitations of the fabrication craft, the slit width and period length must satisfy the relationship of $d,(p-d)\geq2$~mm. Consequently, the driving pulse designed in our work is a compromise consideration. The metamaterial proposed in our work is a binary design, where only two phase responses, $0$ and $\pi$, are required to achieve a desirable transmission field pattern. By utilizing the binary design, other fascinating beam-shaping behaviors such as focusing beam~\cite{WLi2020} and Airy beam~\cite{STang2021} can be realized in addition to beam splitting. However, different from the construction of beam splitter by periodically arraying the acoustic coupler, the design of focusing beam or Airy beam generators requires more complicated phase arrangement of 0 and $\pi$. As a result, phase shifters are required to realize a focusing or Airy beam, which will unavoidably influence the unidirection feature of devices.

	Finally, we have demonstrated a new approach for designing UDAMs, which broadens the applications of NHQTs into the field of classical waves. Three-waveguide couplers can track the evolution of NHQTs by varying the widths of their coupling slits.
	By arraying several three-waveguide couplers of a Hadamard-like transformation, a UDAM is achieved, which can efficiently split a plane acoustic wave in a relatively wide space. The excellent agreement among results of analytical predictions, numerical simulations, and experimental measurements confirms the great applicability of NHQTs in engineering UDAMs. The current work would be a significant addition to the achievements of studying NHQTs and could pave the way for a new method of designing UDAMs with potential applications and excellent properties.\\
	\\
	\begin{acknowledgements}
		\begin{small}{\it The authors acknowledge funding from the  National Natural Science Foundation of China (NSFC) (11675046, 21973023, 11804308); Program for Innovation Research of Science in Harbin Institute of Technology (A201412); Postdoctoral Scientific Research Developmental Fund of Heilongjiang Province (LBH-Q15060); and Natural Science Foundation of Henan Province under Grant No.~202300410481.
				We thank the HPC Studio at School of Physics of Harbin Institute of Technology 37 for access to computing resources through INSPUR-HPC@PHY.HIT.EDU.}\end{small}
	\end{acknowledgements}
	\maketitle
	
	\bibliography{apssamp}
	
\end{document}